\documentclass{tMPH2e}
\usepackage{float}

\begin{document}

\doi{10.1080/0026897YYxxxxxxxx}
\issn{13623028}
\issnp{00268976}
\jvol{00}
\jnum{00} \jyear{0000} 

\articletype{RESEARCH ARTICLE}

\title{{ Density and bond-orientational relaxations in supercooled water}}

\author{
Jeremy C. Palmer$^{a}$$^{\ast}$\thanks{$^\ast$Corresponding author. Email: jcpalmer@uh.edu}, 
Rakesh S. Singh$^{b}$, 
Renjie Chen$^{a}$, 
Fausto Martelli$^{c}$, 
and Pablo G. Debenedetti$^{b}$$^{\dag}$\thanks{$^\dag$Corresponding author. Email: pdebene@princeton.edu}
\\\vspace{6pt} 
$^{a}${\em{Department of Chemical and Biomolecular Engineering, University of Houston, Houston, TX, USA}};
$^{b}${\em{Department of Chemical and Biological Engineering, Princeton University, Princeton, NJ, USA}}
$^{c}${\em{Department of Chemistry, Princeton University, Princeton, NJ, USA}}
\\\vspace{6pt}
}

\maketitle

\begin{abstract}
Recent computational studies have reported evidence of a metastable liquid-liquid phase transition (LLPT) in molecular models of water under deeply supercooled conditions. A competing hypothesis suggests, however, that non-equilibrium artifacts associated with coarsening of the stable crystal phase have been mistaken for an LLPT in these models. Such artifacts are posited to arise due to a separation of time scales in which density fluctuations in the supercooled liquid relax orders of magnitude faster than those associated with bond-orientational order. Here, we use molecular simulation to investigate the relaxation of density and bond-orientational fluctuations in three molecular models of water (ST2, TIP5P and TIP4P/2005) in the vicinity of their reported LLPT. For each model, we find that density is the slowly relaxing variable under such conditions. We also observe similar behavior in the coarse-grained mW model of water. Our findings therefore challenge the key physical assumption underlying the competing hypothesis.
\end{abstract}

\section{Introduction}

\par The existence of a metastable liquid-liquid phase transition (LLPT) in deeply supercooled water was posited by Poole et al.\cite{Poole92} more than two decades ago on the basis of numerical evidence from molecular dynamics (MD) simulations of the ST2 water model \cite{Stillinger74}. The preponderance of evidence from subsequent computational studies supports this hypothesis \cite{Liu09,Sciortino11,Liu12,Kesselring13,Poole13}, suggesting that ST2 phase-separates into a high-density liquid (HDL) and a low-density liquid (LDL) at sufficiently low temperatures. An alternative hypothesis \cite{Limmer13} argues, however, that reported signatures of the LLPT in ST2 and other water models \cite{Mahoney00, Abascal05} are non-equilibrium artifacts associated with crystallization. According to the \emph{artificial polyamorphism} hypothesis (APH) \cite{Limmer13}, all previous reports of LLPT-like behavior occur at temperatures near the stability limit of the liquid $T_s(P)$, in a regime where crystallization occurs by a sluggish coarsening process. Large density fluctuations during coarsening give rise to the illusion of an LDL-like phase with a density similar to that of the stable crystal. This behavior would then arise from a purported separation of relaxation time scales in which fluctuations in density relax rapidly compared to those associated with bond-orientational order. Thus, the APH argues that the LDL is a transient manifestation of the burgeoning ice phase that has equilibrated its density, but not its bond-orientational order \cite{Limmer13}. Although the APH is supported by some free energy calculations \cite{Limmer13} for the ST2 model showing evidence of only a single liquid phase in approximate coexistence with cubic ice near the reported LLPT, these findings are in direct conflict with recent computational investigations \cite{Liu09,Sciortino11,Liu12,Kesselring13,Poole13, Palmer14, Smallenburg15}, and they have not been independently verified.

\par The APH is inconsistent with findings from our recent study of ST2 water \cite{Palmer14} showing evidence of HDL-LDL in coexistence at deeply supercooled conditions. In particular, the free energy surface in Fig. 1 of Ref. \cite{Palmer14} shows that the LDL persists after relaxing all fluctuations accessible to the system by sampling reversibly \emph{back and forth} between the liquid and crystal regions. This is in direct contradiction to the predictions of the APH. The salient features of our free energy results have been independently verified by several research groups \cite{Sciortino11, Poole13, Kesselring13}, and we have carefully scrutinized their accuracy using six different computational techniques \cite{Palmer14}. Thermodynamic analysis shows that the free energy difference between the liquids and the stable ice phase predicted by our simulations is consistent with expectations based on an accurate empirical equation of state for real water \cite{Palmer14}. The LLPT in ST2 also exhibits scaling behavior consistent with a first-order transition over the range of system sizes whose free energy surfaces can be calculated with currently available computational resources \cite{Palmer14}. Furthermore, Smallenburg and Sciortino \cite{Smallenburg15} recently demonstrated that by adjusting a single model parameter, the bond flexibility, the LLPT in ST2 can be traced in a continuous fashion to a thermodynamically stable liquid-liquid transition. Their findings therefore exclude the possibility that crystallization could be mistaken for an LLPT and confirm the reversible nature of our calculations \cite{Palmer14}. Recent work by these authors also shows that LLPTs are an inherent property of fluids with interpenetrating tetrahedral networks similar to those found in ST2 water \cite{Smallenburg14}.

\par Our study \cite{Palmer14} shows that signatures of the LLPT in ST2 are reversible and do not slowly disappear over time or vanish when sampling is performed to and from the crystal region, as incorrectly predicted by the APH \cite{Limmer13}. We are, however, unaware of computational studies that have systematically examined the physical assumption underlying this hypothesis, namely that density fluctuations relax rapidly compared to those associated with bond-orientational order. Here, we use molecular simulation to characterize the relaxation behavior of four different water models: ST2, TIP5P, TIP4P/2005 and mW. Following previous studies \cite{Limmer15, Palmer14}, we characterize the relaxation of fluctuations by computing autocorrelation functions ($C(t)$, ACFs) for the molecular density of the system, $\rho$, and the Nelson-Steinhardt-Ronchetti bond-orientational order parameter, $Q_{6}$ \cite{Steinhardt83}. The parameter $Q_{6}$ quantifies the extent of bond-orientational coherence in the system, and it assumes typical values  $< 0.1$ and $> 0.4$ for configurations with liquid- and crystal-like symmetry, respectively. According to the APH \cite{Limmer13}, $\tau_{Q_6} \gg \tau_{\rho}$ in deeply supercooled water, where $\tau \equiv \int_{0}^{\infty} C(t) \,dt$ is the characteristic relaxation time computed by integrating over the ACF. For each water model, in contrast, we observe that $\tau_{\rho} \geq \tau_{Q_6}$, which suggests the APH is fundamentally inconsistent with the relaxation dynamics of deeply supercooled water. 

\section{Results}

\par
\begin{figure}[H]
\centering
\includegraphics[scale=0.8]{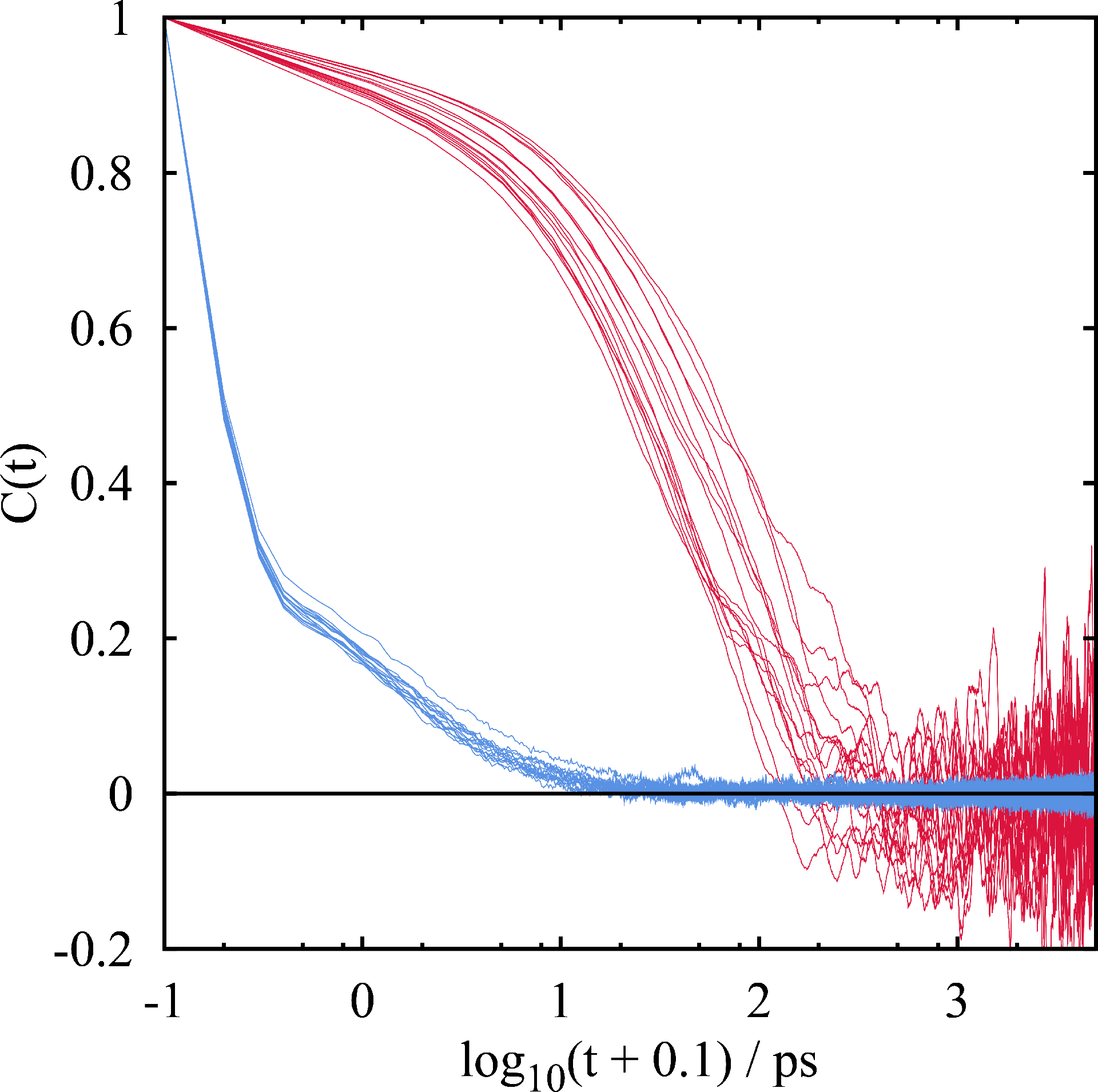}
\includegraphics[scale=0.8]{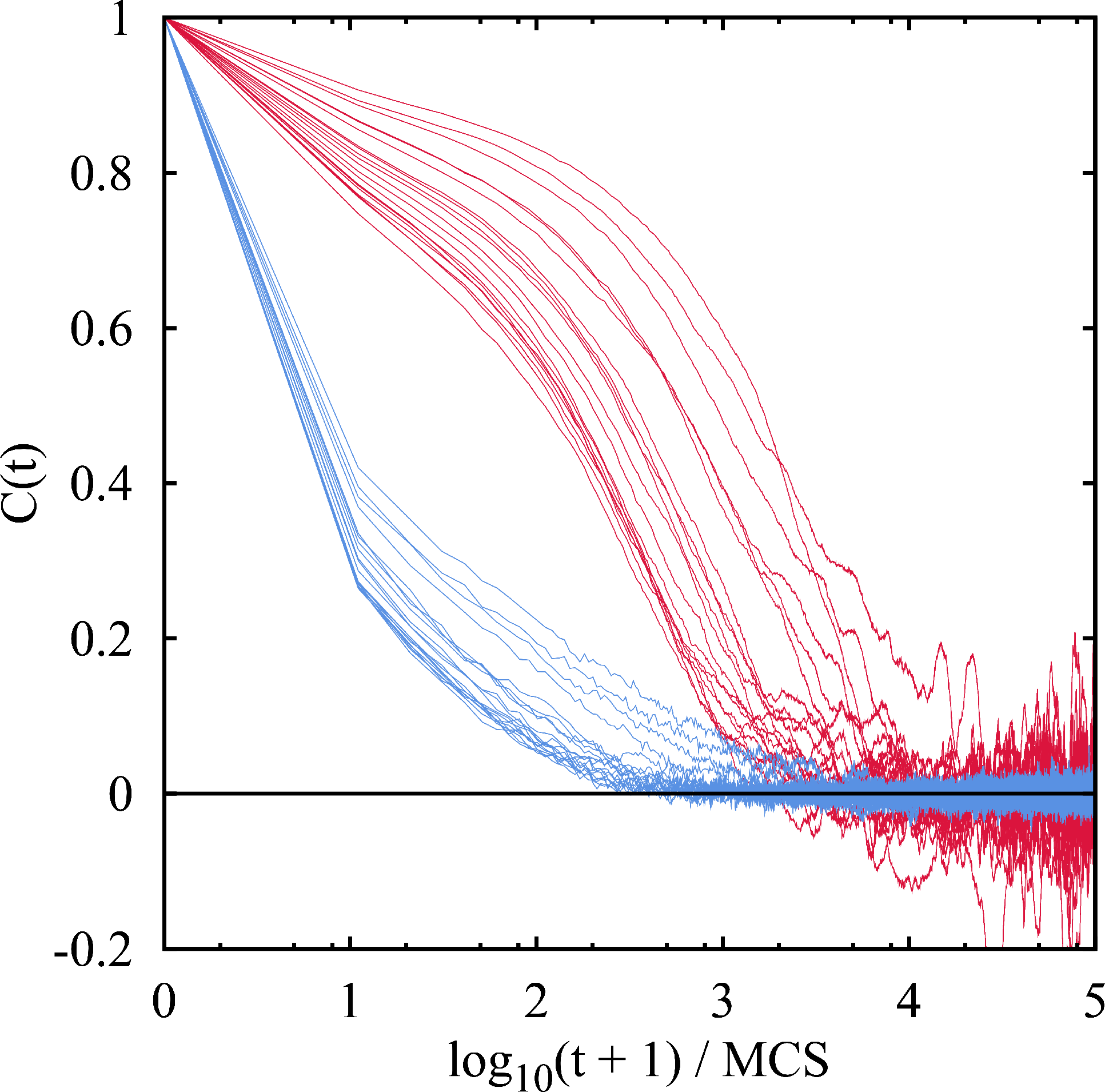}
\caption{ Autocorrelation functions for $\rho$ (red) and $Q_{6}$ (blue) in the HDL region of a system containing 216 ST2-EW(v)\textsuperscript{*} water molecules at 228.6 K and 2.4 kbar, which is the point of liquid-liquid coexistence shown in Fig. 1 of Ref. \cite{Palmer14} . The left panel shows correlation functions computed from 18 representative unrestrained NPT MD simulations, whereas the right panel shows the same functions computed from 20 independent US-HMC simulations restrained at different densities spanning the HDL region. Time is reported in Monte Carlo Sweeps (MCS) for the US-HMC simulations using the same definition of MCS employed in Ref \cite{Palmer14}.}
\label{figure1: fig1}
\end{figure}

 We first examine ST2, which is the only water model where an LLPT has been identified using rigorous free energy techniques designed to investigate phase transitions. There are several variants of ST2 that differ in their treatment of long-range electrostatic interactions. The Ewald variant described in our recent study \cite{Palmer14}, namely one involving the Ewald summation and vacuum boundary conditions, henceforth denoted ST2-EW(v)\textsuperscript{*}, is identical to the model employed by Liu et al. \cite{Liu09,Liu12} (ST2-EW(v)), except that long-range corrections to the Lennard-Jones interactions are neglected. Our previous calculations with this model demonstrate that density and bond-orientational relaxations are highly coupled in the LDL region \cite{Palmer14}. The APH, however, is concerned with relaxation behavior at higher densities \cite{Limmer15} near the HDL basin reported in our study \cite{Palmer14}. This is because the putative physical picture underlying the APH is that of rapid density fluctuations in HDL causing the transient appearance of LDL, which subsequently and slowly gives rise to the stable crystal phase via slow equilibration of $Q_{6}$. By contrast, unrestrained NPT MD simulations of ST2-EW(v)\textsuperscript{*} reveal that fluctuations in $\rho$ relax considerably more slowly than those in $Q_{6}$ in the HDL at 228.6 K and 2.4 kbar (Figure \ref{figure1: fig1}), which is the same condition where we previously demonstrated metastable phase separation in this model \cite{Palmer14}. The system remains liquid-like over the course of the MD simulations as evidenced by the fact that $\left<Q_{6}\right>\approx0.05$. Although some statistical noise is observed in the tails of the ACFs due to sampling over a finite period of time, they fluctuate about zero at long times, demonstrating that they are well-converged. Such convergence is possible under these conditions because the barrier separating HDL and LDL is ca. $4k_{B}T$ \cite{Palmer14}. As a result, MD simulations are able to equilibrate the full range of $\rho$ and $Q_{6}$ fluctuations relevant to the HDL (ca. $1.03 –- 1.25$ g cm\textsuperscript{-3} and $0.01 –- 0.1$, respectively) before the system transitions into LDL. 

\par We also examine the behavior of $\rho$ and $Q_{6}$ ACFs computed using the umbrella sampling hybrid Monte Carlo (US-HMC) scheme described in Ref. \cite{Limmer13} (Figure \ref{figure1: fig1}). The US-HMC method has been previously used to perform free energy calculations for ST2 and investigate its relaxation behavior \cite{Palmer14, Limmer13, Limmer15}. Independent simulations were performed using weak harmonic umbrella restraints to sample twenty different $\rho$ windows spanning the HDL region. The parameter $Q_{6}$ was also restrained to a window centered at 0.05. In accord with our unbiased MD simulations, the ACFs computed with US-HMC show that $\tau_{\rho} > \tau_{Q_6}$.

\par The fact that the qualitative behavior of the system is unaffected by the umbrella restraints or the stochastic nature of the HMC algorithm allows us to use US-HMC to characterize $\rho$ and $Q_{6}$ relaxations in ST2 even under conditions where the HDL and LDL are not separated by a large free energy barrier. As a result, we have used US-HMC to compute $\tau_{\rho}$ and $\tau_{Q_6}$ in the HDL region for the ST2-EW(v) model at all points of liquid-liquid coexistence identified by Liu et al.\cite{Liu09,Liu12} (Table \ref{table1}). Twenty independent US-HMC simulations were performed in the HDL region at each state point. A similar set of calculations was also performed for the reaction-field variant of ST2 (ST2-RF) at each of the points of liquid-liquid coexistence identified by Poole et al. \cite{ Poole13} (Table \ref{table1}). We observe that $\tau_{\rho} > \tau_{Q_6}$ in the HDL over the full range of conditions where LLPT behavior has been reported in both variants of ST2.

\begin{table}[h!]
\tbl{Relaxation times in the HDL region of the ST2 model}
{\begin{tabular}{ccccccc}
\toprule

Model &
$T$ (K)&$P$ (kbar)&
$\left<Q_6\right>^{\rm a}$ &
$\left<\rho\right>^{\rm a}$ (g cm\textsuperscript{-3}) &
$\tau_{\rho}/\tau_{Q_6}^{\rm b}$ \\
\colrule

ST2-EW(v) &
224 &
2.3 &
0.05 &
1.03 -- 1.19 &
\phantom{0}4.6 -- 33.4 \\

&
228 &
2.2 &
0.05 &
1.04 -- 1.18 &
\phantom{0}9.8 -- 28.3 & \\

&
235 &
2.0 &
0.05 &
1.03 -- 1.17 &
12.5 -- 30.2 & \\

&
238 &
1.9 &
0.05 &
1.03 -- 1.16 &
10.2 -- 28.4 & \\

&
240 &
1.8 &
0.05 &
1.03 -- 1.16 &
12.8 -- 36.3 & \\

&
242 &
1.8 &
0.05 &
1.03 -- 1.16 &
\phantom{0}9.4 -- 25.1 & \\

ST2-RF &
230 &
2.45 &
0.05 &
0.95 -- 1.16 &
\phantom{0}5.4 -- 37.5 \\

&
235 &
2.25 &
0.05 &
0.95 -- 1.15 &
12.2 -- 42.5 & \\

&
240 &
2.05 &
0.05 &
0.95 -- 1.14 &
\phantom{0}4.9 -- 55.4 & \\

&
245 &
1.85 &
0.05 &
0.94 -- 1.13 &
\phantom{0}5.1 -- 30.2 & \\

\botrule

\end{tabular}}
\tabnote{$^{\rm a}$Range of average values in umbrella sampling windows}
\tabnote{$^{\rm  b}$Integrated autocorrelation time, $\tau \equiv \int_{0}^{\infty} C(t) \,dt$}
\label{table1}
\end{table}

\par
\begin{figure}[h!]
\centering
\includegraphics[scale=0.8]{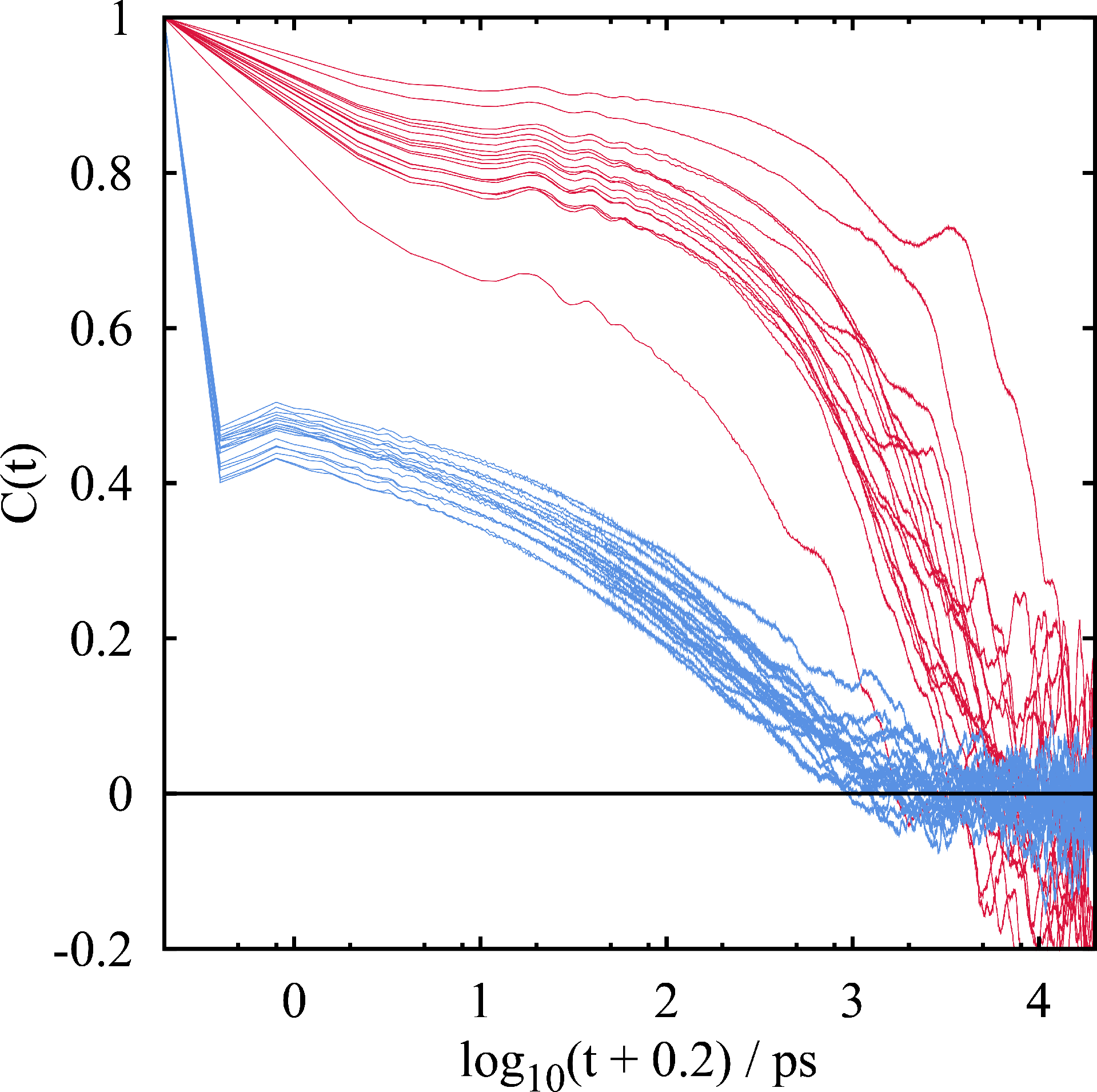}
\includegraphics[scale=0.8]{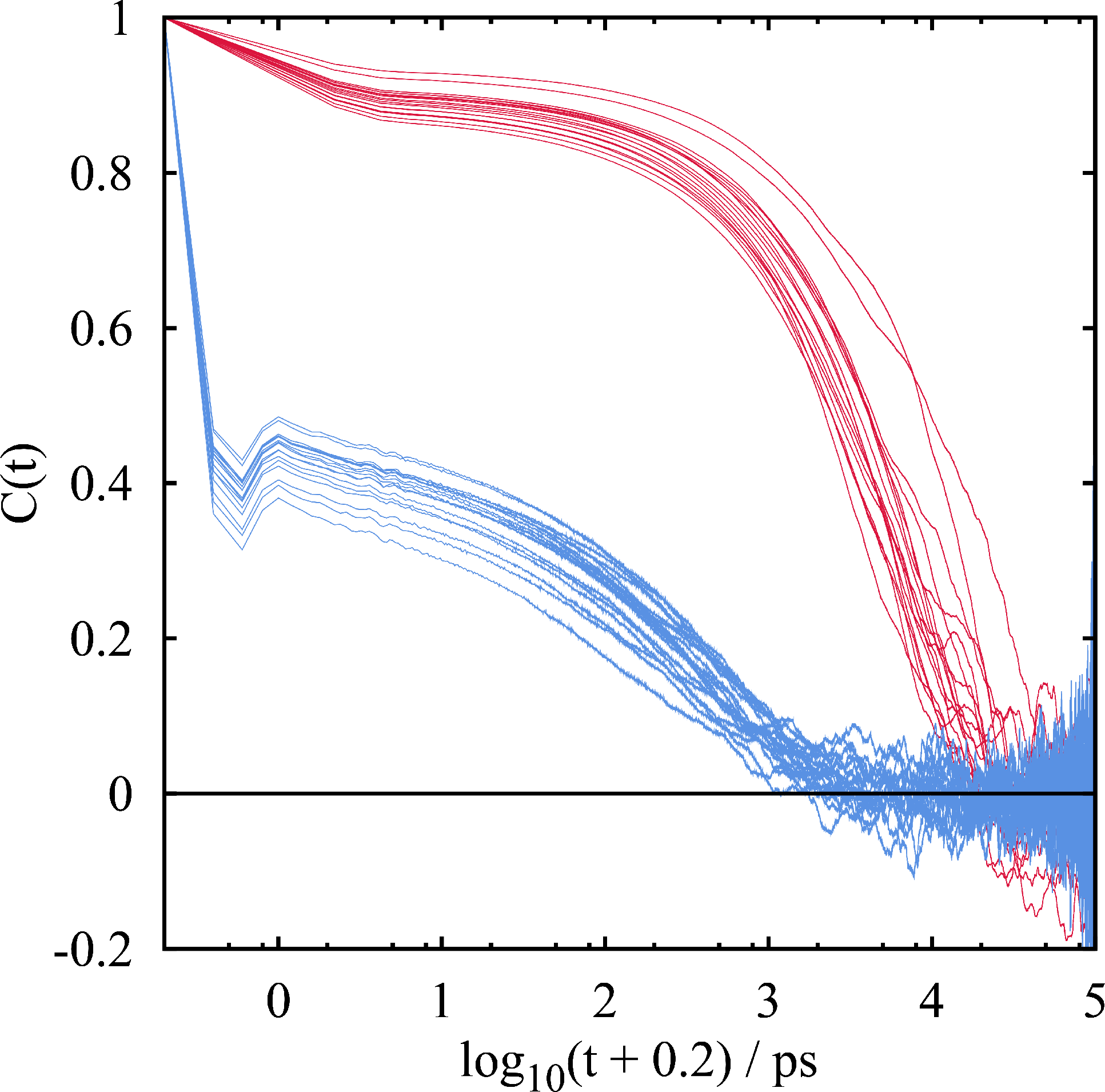}
\caption{ Autocorrelation functions for $\rho$ (red) and $Q_{6}$ (blue) from unrestrained NPT MD simulations of TIP5P at 207 K and 3.6 kbar (left) and TIP4P/2005 at 185 K and 1.8 kbar (right). Qualitatively similar relaxation behavior is also observed in each model at nearby state conditions. The simulated systems contain 512 water molecules.} 
\label{figure2: fig2}
\end{figure}

 Our studies collectively demonstrate that the APH is inconsistent with the relaxation dynamics and reversible phase behavior of ST2 under conditions relevant to its LLPT. Our findings do not exclude the possibility, however, that rapid density fluctuations have been mistaken for LLPT-like behavior in other models such as TIP5P \cite{Mahoney00} and TIP4P/2005 \cite{Abascal05}. To this end, we performed unrestrained MD simulations of these two models. Simulations for TIP5P were conducted at 207 K and 3.6 kbar. We estimate that this is near liquid-liquid coexistence based on the equation of state data reported by Paschek \cite {Paschek05}, which predicts that TIP5P's  liquid-liquid critical point is located at 210 K and 3.1 kbar. Literature estimates of the liquid-liquid critical temperature and pressure for TIP4P/2005 \cite{Abascal10,Bresme14, Sumi13, Singh16} vary between 193 and 182 K and 1.7 and 1.4 kbar, respectively. As a result, simulations for TIP4P/2005 were performed at 185 K and 1.8 kbar. In agreement with our results for ST2, the ACFs calculated at these conditions (Figure \ref{figure2: fig2}) reveal that density is the slowly relaxing variable in both models. Similar behavior is also observed at nearby state conditions ($205-215$ K, $3.2-3.6$ kbar for TIP5P, and $185-205$ K, $1.45-1.8$ kbar for TIP4P/2005). In each case, the systems remain liquid-like over the course of the $\mu$s-long simulations and show no signs of crystallization.

\begin{figure}[h!]
\centering
\includegraphics[scale=1.0]{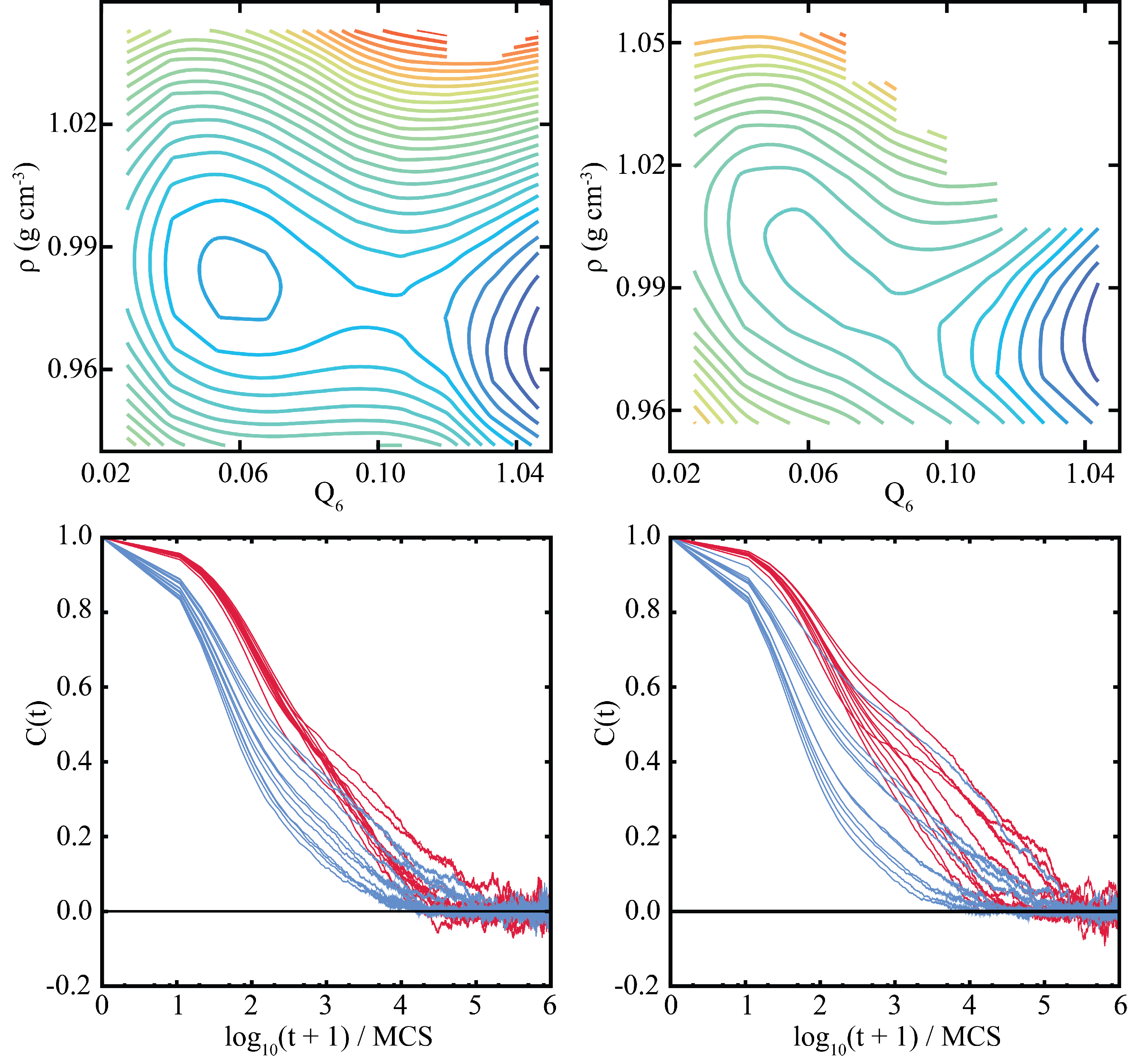}
\caption{ Free energy surfaces and autocorrelation functions computed using US-HMC for a system with 216 mW water molecules. The left and right columns show data at (205 K, 1 bar) and (193 K, 2.3 kbar), respectively. The liquid basin located at $\left<Q_{6}\right>\approx0.05$ vanishes as the conditions are changed from 205 K and 1 bar to 193 K and 2.3 kbar, indicating that the stability limit has been crossed. Autocorrelation functions computed for $\rho$ (red) and $Q_{6}$ (blue) from sampling windows with $\left<Q_{6}\right>\approx0.05$ show that $\tau_{\rho} \geq \tau_{Q_6}$ at both conditions.  The ratio $\tau_{\rho}$ / $\tau_{Q_6}$ in different sampling windows ranges from 1.0 to 3.2 and 1.3 to 7.6 at 205 K and 190 K, respectively. Free energy contours are $1k_{B}T$ apart.} 
\label{figure3: fig3}
\end{figure}

\par The APH states that $\tau_{Q_6} \gg \tau_{\rho}$ at temperatures near the stability limit of the liquid with respect to the crystal $T_s(P)$, which is purportedly the same region where signatures of LLPT behavior have been observed \cite{Limmer13}.  Accordingly, we have examined ST2, TIP5P and TIP4P/2005 at such conditions. Free energy calculations for ST2 unambiguously show that HDL and LDL are metastable with respect to crystallization at liquid-liquid coexistence \cite{Palmer14}, and our analysis of $\rho$ and $Q_{6}$ relaxation dynamics demonstrates that $\tau_{\rho} > \tau_{Q_6}$ under these conditions (Figure \ref{figure1: fig1}, Table \ref{table1}). TIP5P and TIP4P/2005 exhibit similar relaxation dynamics (Figure \ref{figure2: fig2}), and the fact that no signs of crystallization are observed during our MD simulations suggests that conditions relevant to the reported LLPT in these models lie above $T_s(P)$.  To carefully scrutinize the APH, however, we have also performed US-HMC simulations of the coarse-grained mW water model \cite{Molinero09} near $T_s(P)$. Although mW does not exhibit an LLPT \cite{Moore11}, the relaxation behavior assumed by the APH is purportedly valid for all water models \cite{Limmer13_2}. The $\rho - Q_{6}$ free energy surface computed with US-HMC demonstrates that mW is near $T_s(P)$ at 205 K and 1 bar, with only a ca. $1k_{B}T$ barrier to crystallization (Figure \ref{figure3: fig3}).  At higher pressures, $T_s(P)$ decreases, and we observe that the liquid becomes unstable to with crystallization at 190 K and 2.3 kbar. The ACFs obtained from umbrella windows in the liquid region (Figure \ref{figure3: fig3}) show, however, that density is the slowly relaxing variable at both conditions, with $\tau_{\rho}$ / $\tau_{Q_6}$ for different sampling windows ranging from 1.0 to 3.2 and 1.3 to 7.6 at 205 K and 190 K, respectively. In contrast with the underlying assumption of the APH, in this case we find fluctuations in $\rho$ and $Q_{6}$ relax on comparable time scales, with $\tau_{\rho}$ slightly larger than $\tau_{Q_6}$. Although this behavior is different than the one exhibited by the molecular models discussed above (Figures \ref{figure1: fig1}, \ref{figure2: fig2}), the putative inequality $\tau_{Q_6} \gg \tau_{\rho}$ assumed by the APH is once again at odds with the system's actual behavior.

\section{Conclusions}

\par In summary, the APH posits that a sluggish crystallization process has been misinterpreted as an LLPT in models of water. The key physical argument behind this hypothesis is a purported separation of density and bond-orientational relaxation time scales in which $\tau_{Q_6} \gg \tau_{\rho}$ near the stability limit of the liquid. Such behavior has been described as ``self-evident'' \cite{Limmer15}, and it has been argued to be universal for water models \cite{Limmer13, Limmer13_2, Limmer15}. By contrast, our results show that $\tau_{\rho} \geq \tau_{Q_6}$ in this regime for four different models. Recent computational studies have provided unambiguous numerical evidence supporting the existence of an LLPT in ST2 water. Our exhaustive investigations collectively show that at all state conditions relevant to ST2's LLPT, the relaxation dynamics are fundamentally inconsistent with the behavior assumed by the APH. We also observe that density is the slowly relaxing variable near the reported LLPT in TIP5P and TIP4P/2005. Although our results do not answer the question of whether these models exhibit an LLPT, they suggest that the reported signatures of such a transition are not associated with relaxation processes where $\tau_{Q_6} \gg \tau_{\rho}$. We anticipate that future studies using free energy methods will be necessary to characterize the reversible phase behavior of these models and unambiguously identify conditions, if any, where metastable liquid-liquid phase separation occurs. Finally, we have also studied the behavior of the mW model, which does not exhibit an LLPT. Our calculations demonstrate that $\tau_{\rho} \geq \tau_{Q_6}$ in mW, even when the liquid becomes unstable with respect to crystallization. After investigating the behavior of four different water models, we therefore find no evidence to support the APH or its underlying physical assumption. 

\section*{Acknowledgment{s}}{JCP gratefully acknowledges support from the Welch Foundation (Grant E-1882). PGD gratefully acknowledges the support of the National Science Foundation (Grants No. CHE-1213343 and CBET-1263565). FM acknowledges support from the US Department of Energy (Grant No. DE-SC0008626).}

\end{document}